\documentclass[useAMS,usenatbib]{mn2e}
\usepackage{aecompl}
\usepackage{bethmacros}
\usepackage{booktabs}
\usepackage[T1]{fontenc}
\usepackage{graphicx}
\usepackage{xspace}

\def\spose#1{\hbox to 0pt{#1\hss}}
\def\lta{\mathrel{\spose{\lower 3pt\hbox{$\mathchar"218$}}
     \raise 2.0pt\hbox{$\mathchar"13C$}}}
\def\gta{\mathrel{\spose{\lower 3pt\hbox{$\mathchar"218$}}
     \raise 2.0pt\hbox{$\mathchar"13E$}}}
\newcommand\msun{\ensuremath{{\rm M}_\odot}\xspace}
\def\textalpha{\ensuremath{\alpha}\xspace}
\def\wdm{WDM\xspace}
\def\cdm{CDM\xspace}
\def\cf{cf.\xspace}
\def \eg {e.\,g.\xspace}
\def \ie {i.\,e.\xspace}

\def\omegam{\ensuremath{\Omega_{\rm m}}\xspace}
\def\omegab{\ensuremath{\Omega_{\rm b}}\xspace}

\def\omegal{\ensuremath{\Omega_{\rm l}}\xspace}
\def\cesf{\ensuremath{c_{\textrm{\sc esf}}}\xspace}
\newcommand\mwdm{\ensuremath{m_\textrm{\sc wdm}}\xspace}
\newcommand\magicc{MaGICC\xspace}
\newcommand\rtwo{\ensuremath{R_{200}}\xspace}
\newcommand\mtwo{\ensuremath{M_{200}}\xspace}
\newcommand\mgas{\ensuremath{M_{\rm gas}}\xspace}
\newcommand\mstar{\ensuremath{M_\star}\xspace}
\newcommand\mbary{\ensuremath{M_{\rm bary}}\xspace}

\title[MaGICC WDM]{MaGICC-WDM: the effects of warm dark matter in hydrodynamical
simulations of disc galaxy formation}
\author[Herpich et al.]{Jakob Herpich$^1$\thanks{Email: herpich@mpia.de}, Gregory\,S. Stinson$^{1}$, 
Andrea\,V. Macci\`o$^1$, Chris Brook $^{2}$,  
\newauthor{James Wadsley$^{3}$, Hugh M. P. Couchman$^3$, Tom Quinn$^{4}$}
\vspace*{6pt}\\
$^{1}$Max-Planck-Institut f\"ur Astronomie, K\"onigstuhl 17, D-69117 Heidelberg, Germany\\
$^{2}$Departamento de F\'{i}s\'{i}ca Te\'{o}rica, Universidad Aut\'{o}noma de Madrid, E-28049 Cantoblanco, Madrid, Spain\\
$^{3}$Department of Physics and Astronomy, McMaster University, Hamilton, Ontario L8S 4M1, Canada\\
$^{4}$Astronomy Department, University of Washington, Box 351580, Seattle, WA 98195-1580, USA\\
}

\begin{document}
\maketitle

\begin{abstract}
We study the effect of warm dark matter (WDM) on hydrodynamic
simulations of galaxy formation as part of the Making
Galaxies in a Cosmological Context (MaGICC) project.
We simulate three different galaxies using three WDM
candidates of 1, 2 and 5\,keV and compare results with
pure cold dark matter simulations.
WDM slightly reduces star formation and produces less centrally concentrated
stellar profiles. These effects are most evident for the 1\,keV candidate but
almost disappear for $\mwdm>2$\,keV.
All simulations form similar stellar discs independent of
WDM particle mass. In particular, the disc 
scale length does not change when WDM is considered.
The reduced amount of star formation in the case of 1\,keV particles
is due to the effects of WDM on merging satellites
which are on
average less concentrated and less gas rich. 
The altered satellites cause a reduced starburst during mergers because they 
trigger weaker disc instabilities in the main galaxy.
Nevertheless we show that disc galaxy evolution is much more sensitive to stellar feedback
than it is to WDM candidate mass.
Overall we find that WDM, especially when restricted to current
observational constraints ($\mwdm>2$\,keV), has a minor impact on disc galaxy formation.
\end{abstract}


\section{Introduction}
\label{sec:intro}

In the cold dark matter (CDM) framework, dark matter (DM) is composed of massive particles
that interact only gravitationally.
Due to their mass, {cold} dark matter particles have a negligible velocity dispersion
when they decouple from the primordial plasma, {so} they are dynamically cold.

Results of numerical $N$-body simulations of structure formation in the Universe based on
the CDM model are in excellent agreement with observations of the large-scale structure of the
Universe, such as the observed clustering of galaxies
{\citep[\eg][]{Springel2005}.}  

Despite this remarkable success, predictions of the CDM model based on collisionless simulations
seem to contradict observations on galactic and sub-galactic scales.
\cdm-based simulations predict $\sim50$ times as many satellites for Milky Way (MW) mass
DM haloes than can be found orbiting the MW \citep{Moore1999a}.
\citet{Klypin1999} found a similar result for the MW and Andromeda galaxies.
Whereas \cdm predicts that satellite mass functions are self-similar between different halo
masses, observed satellite luminosity functions vary a lot as a function of the mass of
the central object \citep[\eg][]{Diemand2004}.
This discrepancy constitutes the \emph{Missing-Satellites problem}.
Other studies have considered baryonic processes like cosmic reionization
and stellar feedback and found that they 
limit gas cooling into small satellites, making them either too dim to be observed 
or completely dark
\citep[\eg][]{Efstathiou1992,Quinn1996,Bullock2000,Somerville2002,Koposov2009,Okamoto2009,Macci`o2010a,Font2011,Nickerson2011}.

\citet{Boylan-Kolchin2011} showed that not only the number of observed satellites
but also their mass profiles are inconsistent with the collisionless CDM prediction.  The 10 satellites most massive at the time of accretion
in the Aquarius simulations \citep{Springel2008} have rotation
curves that
rise more steeply than the observed rotation velocity of the 10 
most luminous satellites in the MW halo. 
These satellites are `too big to fail'; their analogues 
should form dwarf galaxies sufficiently luminous to be observed in the MW.
It remains unclear whether baryons can alter the DM density profiles enough in such small objects to solve this problem \citep[\eg][]{Zolotov2012,Garrison-Kimmel2013}.

Due to these shortcomings of \cdm, alternative DM particles have been suggested.
One possible particle species is a thermal relic warm dark matter (WDM).
Such WDM particles decouple much earlier than CDM particles
due to their lower mass (keV scale for WDM versus GeV and higher for CDM).
Thus, the WDM particles have a non-negligible velocity dispersion at the time of decoupling.
Due to their velocity dispersion, WDM particles can escape from shallow potential wells and
diffuse into underdense regions so that small-scale density perturbations are reduced instead of  amplified.
The particle diffusion is called \emph{free streaming} and causes the WDM power spectrum to be
suppressed on small scales (i.\,e. large $k$).
As a result, structure formation takes longer in WDM models compared to CDM on all scales,
and the amount of substructure on small scales is suppressed 
\citep[\eg][]{Colin2000,Bode2001,Knebe2002,Gao2007,KuziodeNaray2010,Schneider2012}.
{At the same time surviving substructures have rotation curves in better
agreement with observations \citep{Lovell2012,Anderhalden2013}.}

Thermal relic WDM particles are only one possible
WDM particle family.
There are many other WDM candidates including sterile neutrino
and gravitinos \citep{Hansen2002,Abazajian2006,Boyarsky2009}.
There is a one-to-one relation between sterile neutrino
mass and thermal relic candidate mass, however, so it is always possible to recast constraints for
a thermal candidate in terms of a sterile neutrino mass \citep{Colombi1996}.
Other more complex models have been extensively discussed in the literature, like 
mixed models \citep[\eg][]{Boyarsky2009,Macci`o2013}
or composite DM models \citep{Khlopov2005,Khlopov2008}.

Observational results have ruled out the most extreme WDM candidates.
Early measurements of the Lyman-\textalpha forest in spectra of quasi-stellar objects (QSOs)
implied a lower limit on the mass of WDM thermal relic particles of
$\mwdm \gta 1\,{\rm keV}$ \citep{Narayanan2000,Hansen2002,Viel2005,Viel2006,Viel2008,Seljak2006,Boyarsky2009a}.
Fluxes from gravitationally lensed QSOs \citep{Miranda2007} and the distribution of satellites in the MW \citep{Macci`o2010} provide similar constraints.
The most recent study of high redshift quasars' flux power spectra \citep{Viel2013} tightened
the constraints further to exclude
\wdm candidate masses below $3.3\,{\rm keV}$ at a confidence level of $2\sigma$.

So far most of the studies focusing on the effect of 
WDM on galaxy formation have been based on pure DM simulations
\citep[\eg][and references therein]{Schneider2012} or
combine collisionless simulations with semi-analytical models of galaxy formation
\citep{Macci`o2010,Menci2012,Menci2013,Kang2013}.

So far, though, one of the most powerful tools for studying galaxy formation and
evolution in a cosmological context, high resolution hydrodynamical 
simulations, have yet to consider WDM.
Recently, several groups have been able to form realistic
looking disc galaxies simulated in a \cdm universe over a wide
mass range \citep{Robertson2006,Governato2007,Agertz2011,Guedes2011,
Brook2012a,Scannapieco2012,Marinacci2013,Stinson2013}.

In this study, we aim to fill this gap by combining high resolution hydrodynamical 
simulations with WDM-based cosmological models.
This study is part of the Making Galaxies in a Cosmological Context (MaGICC) project, a 
large simulation campaign aiming to unveil the process of galaxy formation
in a Universe ruled by DM (and dark energy).

Simulations performed within the MaGICC project in a CDM scenario have been 
successful in reproducing many observations of galaxies.
The simulations follow both the $z=0$ and the redshift
evolution of the stellar to halo mass relation \citep[\eg][]{Moster2010,Moster2013} as
detailed in \citet{Stinson2013}.  They reproduce the distribution of metals
(e.g. O {\sc VI}) around star-forming galaxies \citep{Stinson2012} and create 
galaxy rotation curves in agreement with observations \citep{Macci`o2012a,DiCintio2013a}.
They form discs with an old, thick component and a young, thin component that match
the chemical patterns observed in the MW \citep{Stinson2013a}.

In this work, we resimulate three galaxies introduced in the McMaster Unbiased
Galaxy Simulations (MUGS) project \citep{Stinson2010} 
using various thermal
WDM models (5, 2 and 1\,keV) to assess the impact of WDM
on galaxy properties.
The paper is organized as follows. \S \ref{sec:simulations} describes the simulations that 
were used for this comparison.  
\S \ref{sec:results} details the main properties of the simulated galaxies and 
\S \ref{sec:conclusions} presents our conclusions.

\section{Simulations}
\label{sec:simulations}

The three galaxies from the MUGS project are g1536, g5664 and g15784
\citep[for details see][]{Stinson2010}.
The galaxies are simulated in a periodic box $50\,h^{-1}{\rm Mpc}$ on a side
from redshift $z=99$.
The simulation parameters are consistent with a {\it Wilkinson Microwave Anisotropy Probe} 3 cosmology
\citep[$H_0=73\,{\rm km\,s}^{-1}{\rm Mpc}^{-1}$, $\omegam=0.24$, 
$\omegal=0.76$, $\omegab=0.04$, $\sigma_{\rm 8}=0.76$][]{Spergel2007}.
Each galaxy is simulated using three different \wdm
models: $\mwdm=5{\rm\,keV}$ (\wdm{}5), $2{\rm\,keV}$ (\wdm{}2), 
$1{\rm\,keV} $(\wdm{}1) and, for reference, one realization of a 
standard \cdm model.  Initial conditions were created with and without
baryons with identical resolution.
The DM particle mass is $1.1\times10^6\,\msun$, the initial gas particle
mass is $2.2\times10^5\,\msun$ and the initial star particle mass is 
$6.3\times10^4\,\msun$.
All the properties of the different galaxies with baryons are summarized
in Table \ref{tab:galaxyProps}.

\begin{table*}
    \caption[Summary of simulated galaxies]{
        A summary of the properties of simulated galaxies, that includes the virial
        (\mtwo), stellar (\mstar), gas (\mgas) and total baryonic mass (\mbary)
        as well as the virial radius (\rtwo).
        Note that only the results for those simulations including baryons are presented.
    }
    \label{tab:galaxyProps}
    \begin{tabular}{llcccccc}
    \toprule
    Galaxy & Model & \mwdm & \mtwo & \mstar & \mgas & \mbary & \rtwo \\
    & & (keV) & ($10^{10}\,\msun$) & ($10^{10}\,\msun$) & ($10^{10}\,\msun$) & ($10^{10}\,\msun$) & (kpc) \\
    \midrule
    g1536 & CDM & $\infty$ & 56 & 2.4 & 4.9 & 7.3 & 166 \\
    g1536 & WDM5 & 5 & 56 & 2.1 & 4.8 & 6.9 & 165 \\
    g1536 & WDM2 & 2 & 56 & 1.8 & 5.0 & 6.8 & 165 \\
    g1536 & WDM1 & 1 & 55 & 1.5 & 5.0 & 6.5 & 165 \\
    g5664 & CDM & $\infty$ & 44 & 2.7 & 3.1 & 5.8 & 153 \\
    g5664 & WDM5 & 5 & 44 & 2.3 & 3.1 & 5.5 & 152 \\
    g5664 & WDM2 & 2 & 45 & 2.7 & 3.3 & 5.9 & 154 \\
    g5664 & WDM1 & 1 & 39 & 0.7 & 3.9 & 4.5 & 148 \\
    g15784 & CDM & $\infty$ & 121 & 8.3 & 9.5 & 17.8 & 214 \\
    g15784 & WDM5 & 5 & 123 & 7.9 & 9.1 & 17.0 & 215 \\
    g15784 & WDM2 & 2 & 121 & 8.2 & 9.2 & 17.3 & 213 \\
    g15784 & WDM1 & 1 & 119 & 8.9 & 9.0 & 18.0 & 213 \\
    \bottomrule
    \end{tabular}
\end{table*}

\subsection{Initial conditions}
\label{sec:ics}

For the \cdm simulations we used the initial conditions files from the 
MUGS project.
New initial conditions were created for each \wdm model  
from the same region of space as every MUGS galaxy.
The \wdm initial condition creation is identical to the \cdm model 
\citep{Stinson2010} except for a modified initial DM power spectrum.
The same random seed was used for the generation of both warm and \cdm initial conditions.

The \wdm power spectra are computed using a `relative transfer function' to 
the \cdm power spectrum as in \citet{Bode2001}:

\begin{equation}
    P_{\rm WDM}\left(k\right) =
    \left[T_{\rm WDM}\left(k\right)\right]^2 P_{\rm CDM}\left(k\right),
    \label{eq:powertf}
\end{equation}
with
\begin{equation}
    T\left(k\right) = \left[1+\left(\alpha k\right)^{2\nu}\right]^{-5/\nu},
    \label{eq:tf}
\end{equation}
where $\alpha$ is the length-scale of the break in the WDM power spectrum.
\citet{Viel2005} found $\nu=1.12$ and
\begin{equation}
    \alpha=0.049\left(\frac\mwdm{1\,{\rm keV}}\right)^{-1.11}
    \left(\frac{\Omega_{\rm WDM}}{0.025}\right)^{0.11}
    \left(\frac h{0.7}\right)^{1.22}h^{-1}{\rm Mpc}.
    \label{eq:alpha}
\end{equation}

\begin{figure}
\includegraphics[width=\columnwidth]{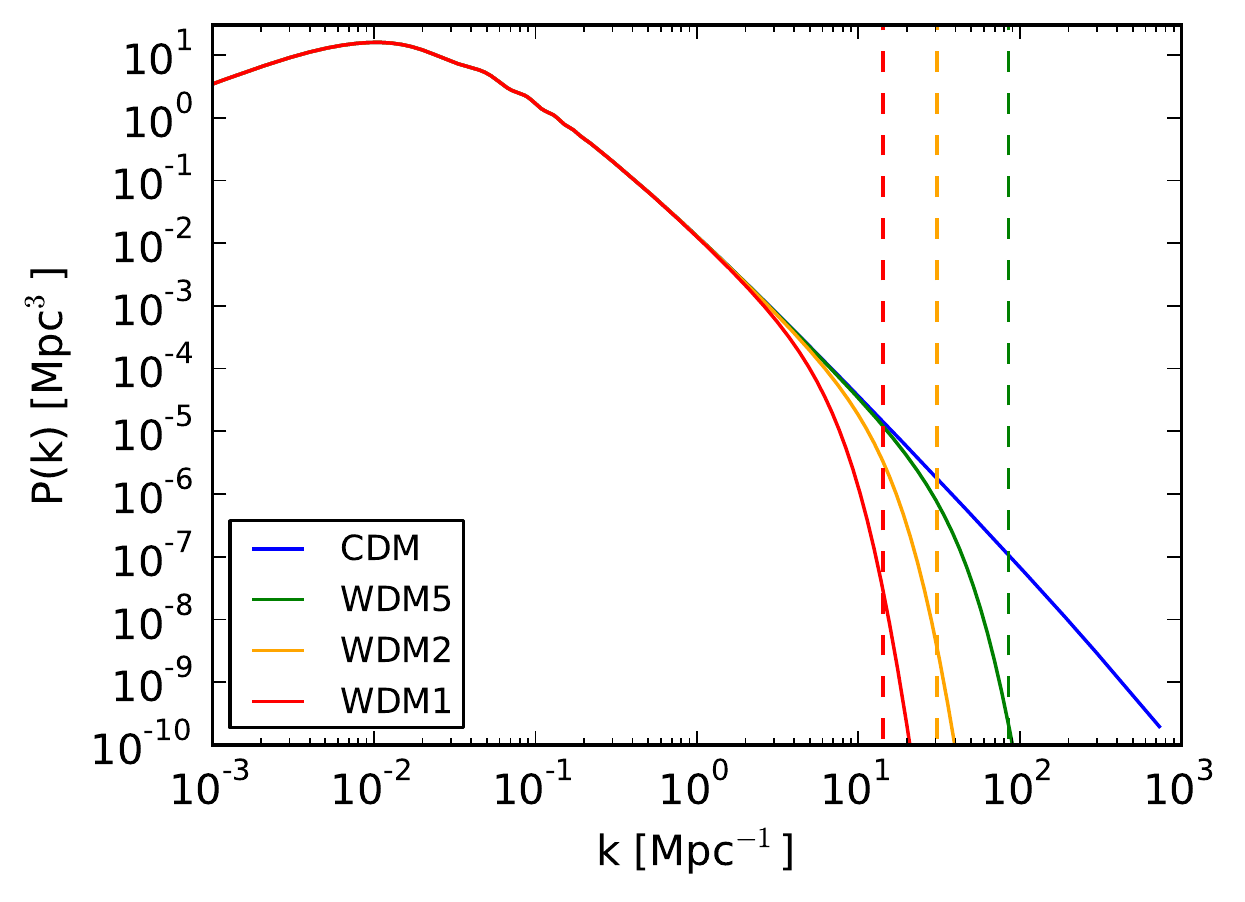}
\caption[WDM power spectra]{
    WDM power spectra.
    The figure shows the power spectra at $z=99$ for all simulated DM
    candidates.
    The blue, green, yellow and red lines correspond to CDM, WDM5, WDM2 and WDM1,
    respectively.
    The vertical dashed lines indicate the scale of the break in the WDM power spectra $\alpha$.
    The figure shows the suppression of power at large $k$ in the WDM
    initial conditions.
    These power spectra are used to sample initial conditions.
}
\label{fig:pspec} 
\end{figure}

Fig. \ref{fig:pspec} shows the initial power spectra at $z=99$ for all simulated
DM candidates.
From the figure, it is evident that power on small spatial scales (large $k$) is suppressed
for warmer DM candidates.
Given the mass resolution of the simulation and the choice of WDM models, 
the contribution from \wdm velocity
dispersion is ignored since it is less than 1\% of the initial Zel'dovich velocity
\citep{Macci`o2012}.
Therefore, only the WDM modification to the power spectrum is considered.

\subsection{Hydrodynamics}
\label{sec:hydrophys}

The simulations use the smoothed particle hydrodynamics code 
\textsc{gasoline} \citep{Wadsley2004}.  It includes metallicity-dependent gas 
cooling, star formation (SF) and a detailed chemical enrichment model.
The physics used in the MaGICC project is detailed in 
\citet{Stinson2013}.  Briefly, stars are formed from gas cooler than 
$T_{max}$ = $1.5\times10^4$ K
and denser than 9.3 cm$^{-3}$ according to the Kennicutt-Schmidt law \citep{Kennicutt1998}
as described in \citet{Stinson2006} with a star formation efficiency (SFE)
parameter $c_\star$=0.1.  The SF density threshold is then set to 
the maximum density at which gravitational instabilities can be resolved, 
$32 M_{gas}\epsilon^{-3}$($n_{th} > 9.3$ cm$^{-3}$), where 
$M_{gas}=2.2\times10^5$ M$_\odot$ and $\epsilon$ is the gravitational 
softening (310 pc).

The star particles are $6.3\times10^{4}$ M$_\odot$, massive
enough to represent an entire stellar population consisting of stars with 
masses given by the \cite{Chabrier2003} initial mass function.  20\% of these 
have masses greater than 8 M$_\odot$ and explode as Type II supernovae (SNeII) from 3.5 
until 35 Myr after the star forms, based on
the Padova stellar lifetimes \citep{Alongi1993, Bressan1993}.  Each supernova (SN)
inputs $E_{SN}=10^{51}$ erg of purely thermal energy into the surrounding gas.  
This energy would be radiated away before it had any dynamical impact because 
of the high density of the star forming gas \citep{Katz1992}.  Thus, the 
SN feedback relies on temporarily delaying cooling
based on the subgrid approximation of a blastwave as described in 
\cite{Stinson2006}.

The SNe feedback does not start until 3.5 Myr after the
first massive star forms.  However, nearby molecular clouds show evidence 
of being blown apart \emph{before} any SNeII have exploded.  
\citet{Pellegrini2007} emphasized the energy input from stellar winds and 
ultraviolett (UV) radiation pressure in M17  prior to any SNeII explosions.
\citet{Lopez2011} found similar energy input into 30 Doradus.  
Thus, in the time period before SNe start exploding, we distribute 
10\% of the luminosity produced in the stellar population, an amount equivalent to the 
UV luminosity, to the surrounding gas without disabling the cooling.
Following the convention of \citet{Stinson2013}, we call this feedback from massive stars:
`Early Stellar Feedback' (ESF).
Most of the energy coming from the ESF is immediately radiated away, but \citet{Stinson2013} have shown
that this ESF has a significant effect on the star 
formation history (SFH) of a MW mass galaxy and places the halo on the 
\cite{Moster2013} stellar mass--halo mass relationship at $z=0$. 

To test how the impact of \wdm compares with the effect of
stellar feedback, we ran an additional suite of simulations of the $\mwdm=1\,{\rm keV}$
initial conditions with a range of reduced ESF efficiencies.
A summary of these galaxies' properties is given in Table \ref{tab:reducedProps}.

\begin{table*}
    \caption[Summary of simulated galaxies]{
        A summary of the properties of simulated representations of g1536 with reduced
        ESF in the \wdm{}1 model, that includes the virial
        (\mtwo), stellar (\mstar), gas (\mgas) and total baryonic mass (\mbary)
        as well as the virial radius (\rtwo).
        Note that only the results for those simulations including baryons are presented.
    }
    \label{tab:reducedProps}
    \begin{tabular}{lcccccc}
    \toprule
    Label & \cesf & \mtwo & \mstar & \mgas & \mbary & \rtwo \\
    & (\%) & ($10^{10}\,\msun$) & ($10^{10}\,\msun$) &
        ($10^{10}\,\msun$) & ($10^{10}\,\msun$) & (kpc) \\
    \midrule
    ESF 10 & 10 & 55 & 1.5 & 5.0 & 6.5 & 165 \\
    ESF 7.5 & 7.5 & 59 & 3.6 & 4.6 & 8.1 & 168 \\
    ESF 5 & 5 & 60 & 4.2 & 4.6 & 8.8 & 170 \\
    ESF 1 & 1 & 59 & 3.4 & 4.7 & 8.2 & 168 \\
    ESF 0 & 0 & 57 & 2.8 & 4.6 & 7.3 & 167 \\
    \bottomrule
    \end{tabular}
\end{table*}

\subsection{Halo identification}
The Amiga Halo Finder \citep{Knollmann2009} identifies all the bound
particles in the simulated galaxies.
The virial overdensity was chosen to be 200, so that the virial radius and 
virial mass are \rtwo and \mtwo, respectively.
The centre for all profile plots was calculated using the shrinking sphere 
algorithm \citep{Power2003}.
To track the evolution of certain properties of the simulated galaxies the 
most massive progenitor at $z=4.5$ has been identified and traced up to $z=0$.

\section{Results}
\label{sec:results}
We first examine the basic properties of the galaxies as a function of
DM candidate mass.  We then re-examine those properties for the WDM1
case, in runs performed with different stellar feedback parametrization.

\subsection{From cold to warm}
\label{sec:wdmsims}
\begin{figure*}
\includegraphics[width=\textwidth]{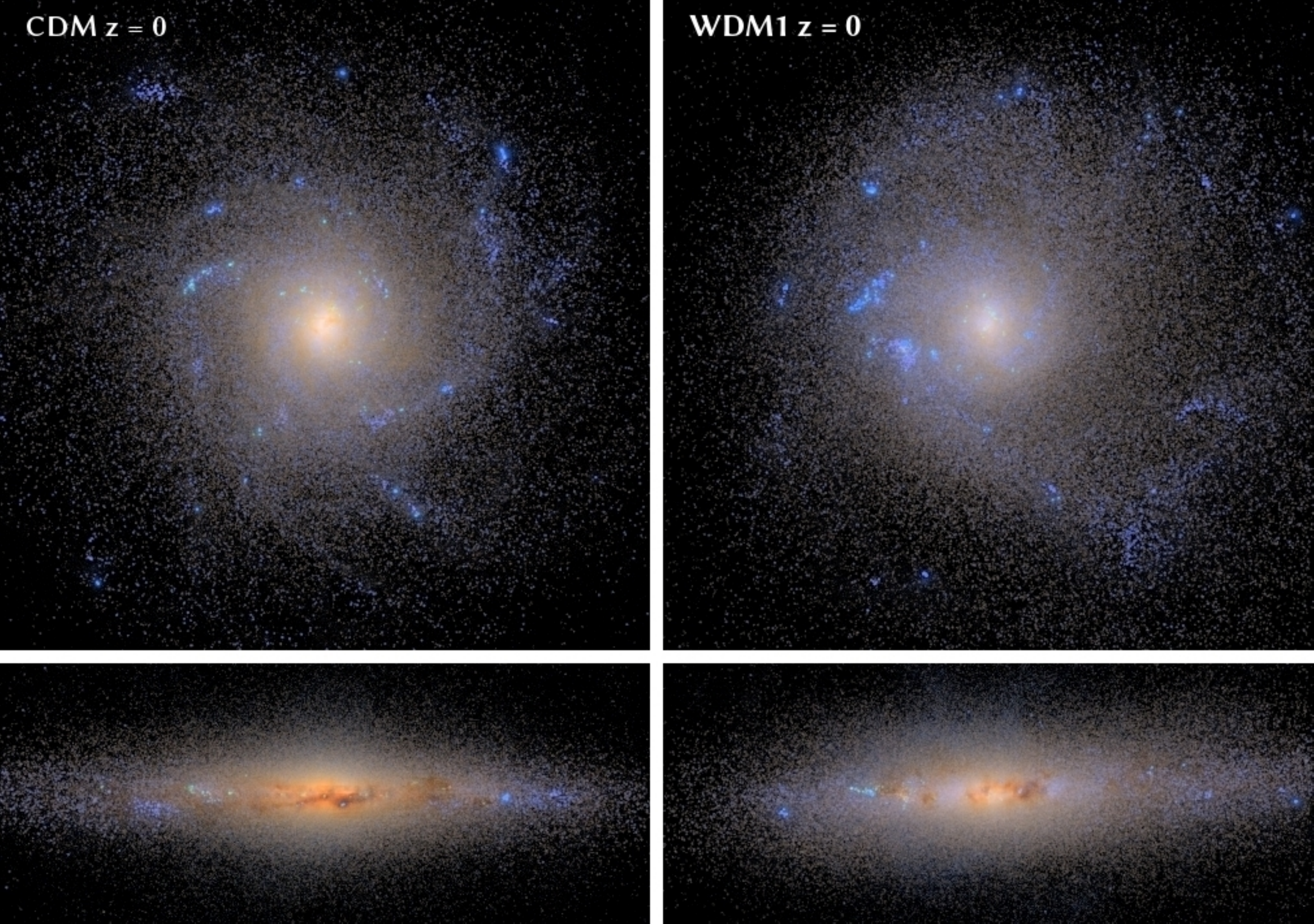}
\caption[Mock observational images]{
        Face-on (upper panel) and edge-on (lower panel) images of the simulated
        galaxy g1536 at $z=0$.
        Each panel is 50\,kpc on a side.
        The left-hand panel shows the CDM simulation while the right-hand panel shows the
        WDM1 simulation.
        The images were created using the Monte Carlo radiative transfer code
        {\sc sunrise}.
        Brightness and contrast in the image are scaled as $a\sin h$ as described
        in \citet{Lupton2004}.
}
\label{fig:sunrise}
\end{figure*}

Fig. \ref{fig:sunrise} shows mock observational images of the simulated galaxy g1536
at $z=0$ in the two extreme DM models: CDM and WDM1.
The images are 50\,kpc on a side.
They were created using the radiative transfer code {\sc sunrise} \citep{Jonsson2006}.
Both images show signs of spiral patterns of young stars (blue) with the CDM simulation
showing perhaps more structured spiral arms.
The WDM1 simulation has a notably reduced stellar density in the centre compared to the CDM
simulation.
{The edge-on view (bottom panels) shows that the} stars are assembled in a disc in both cases.

\begin{figure*}
\includegraphics[width=\textwidth]{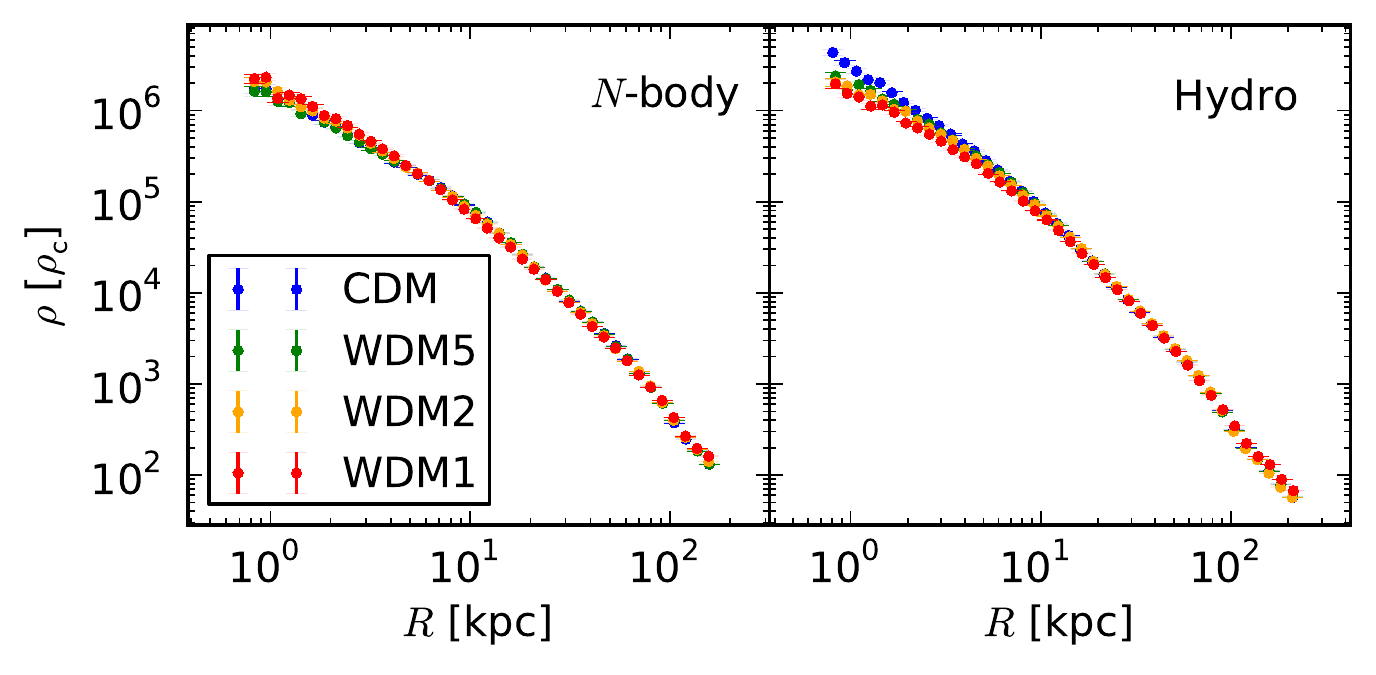}
\caption[DM density profiles]{
    DM density profiles of all simulations of the g1536 galaxy.
    The left-hand panel shows the density profiles of the pure $N$-body simulations, the right-hand
    panel shows the results for the hydrodynamic simulations.
    The blue, green, yellow and red points correspond to the CDM, WDM5, WDM2 and WDM1
    models, respectively.
    The radial range is $2.5\epsilon < R < R_{200}$.
    There is no significant influence of the nature of DM on its density profile
    in $N$-body simulations.
    However, baryonic physics may alter the central DM profile as the right-hand panel
    shows.
}
\label{fig:dmprofs}
\end{figure*}

Fig. \ref{fig:dmprofs} shows DM density profiles of all 
realizations of g1536 {with} (right-hand panel) and without (left-hand
panel) {baryonic physics}.
In the simulations {without baryons}, all DM models yield 
the same DM density profile.
This is consistent with previous studies of the concentration--mass relation 
in WDM models at the scale of massive disc galaxies \citep{Schneider2012}.
In simulations {using baryonic physics,} the DM density profiles 
in the centre are steeper for colder DM models.
{As we will show,} these steeper profiles are due to the dynamical effects of increased SF in the
centre of the \cdm simulations of g1536.  
The results are consistent with the recent study by \citet{DiCintio2013a}, which has shown that
for a galaxy at the peak of the SFE, the effect of baryons is to 
mildly contract the DM density profile.

{The slightly steeper density profiles are also reflected in} the rotation
curves (left-hand panel of Fig. \ref{fig:rcs}) and surface brightness profiles (left-hand panel
of Fig. \ref{fig:sbprofs}) of g1536.
This suggests that WDM has no direct effect on the DM distribution 
of the simulated halo.
Instead, they are a secondary consequence of the effect of WDM on SF in the galactic disc.

\subsubsection{Stellar mass--halo mass relationship}
{In the \magicc\ project, the main constraint is the stellar mass--halo mass
relation \citep[\eg][hereafter called \emph{Moster relation}]{Moster2013}.}
Therefore, we {first} compare all simulated galaxies with {this 
fundamental relationship}.

\begin{figure*}
\includegraphics[width=.5\textwidth]{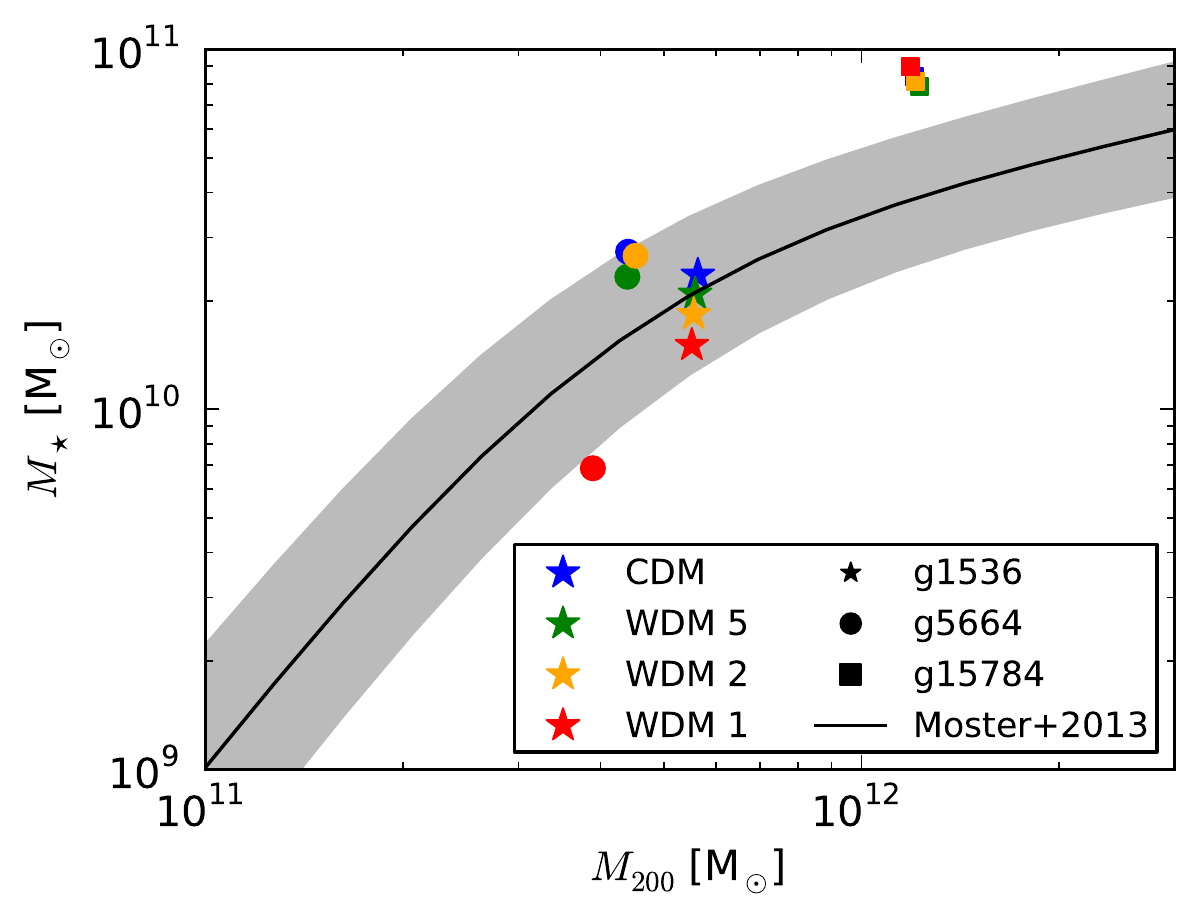}%
\includegraphics[width=.5\textwidth]{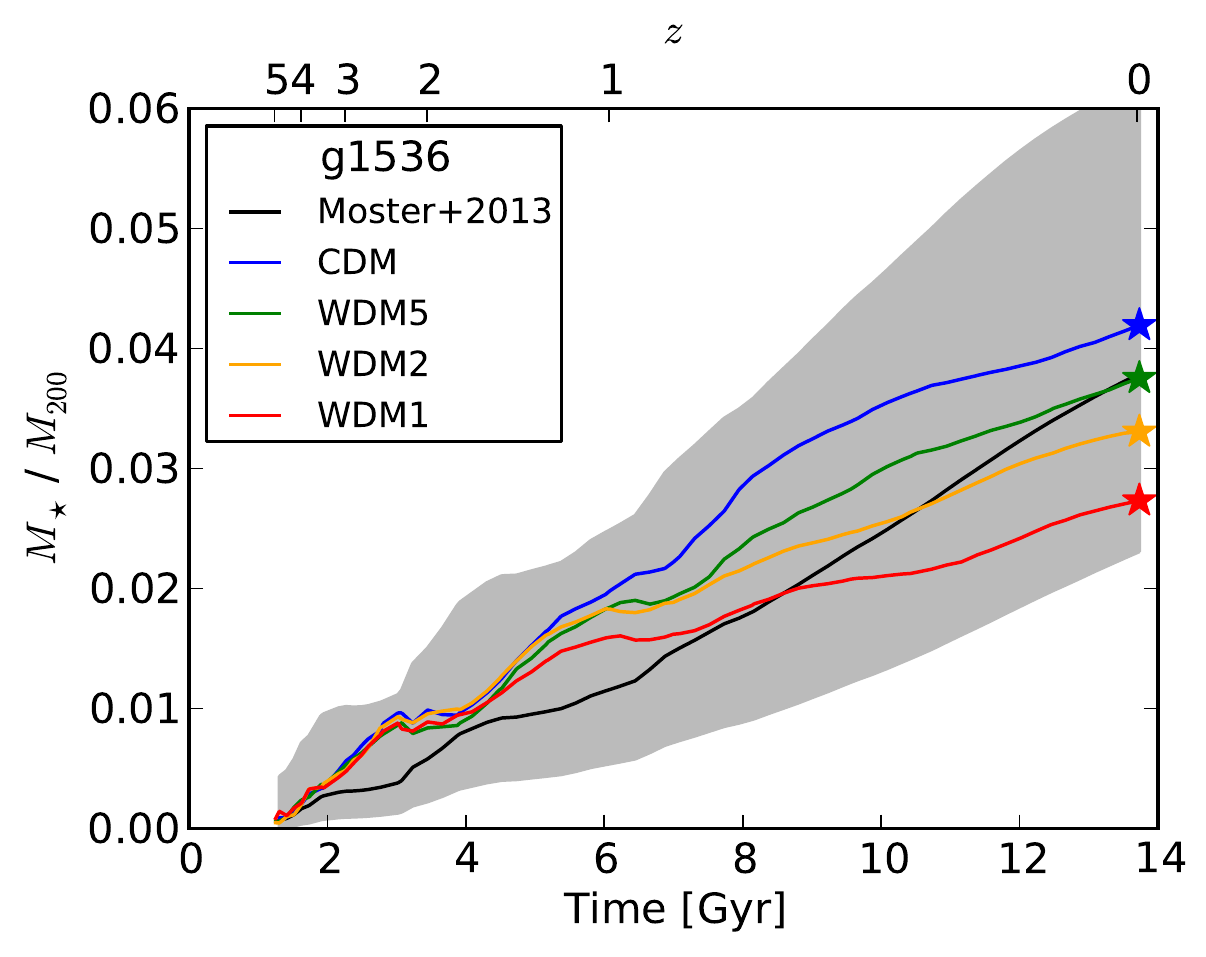}
\includegraphics[width=.5\textwidth]{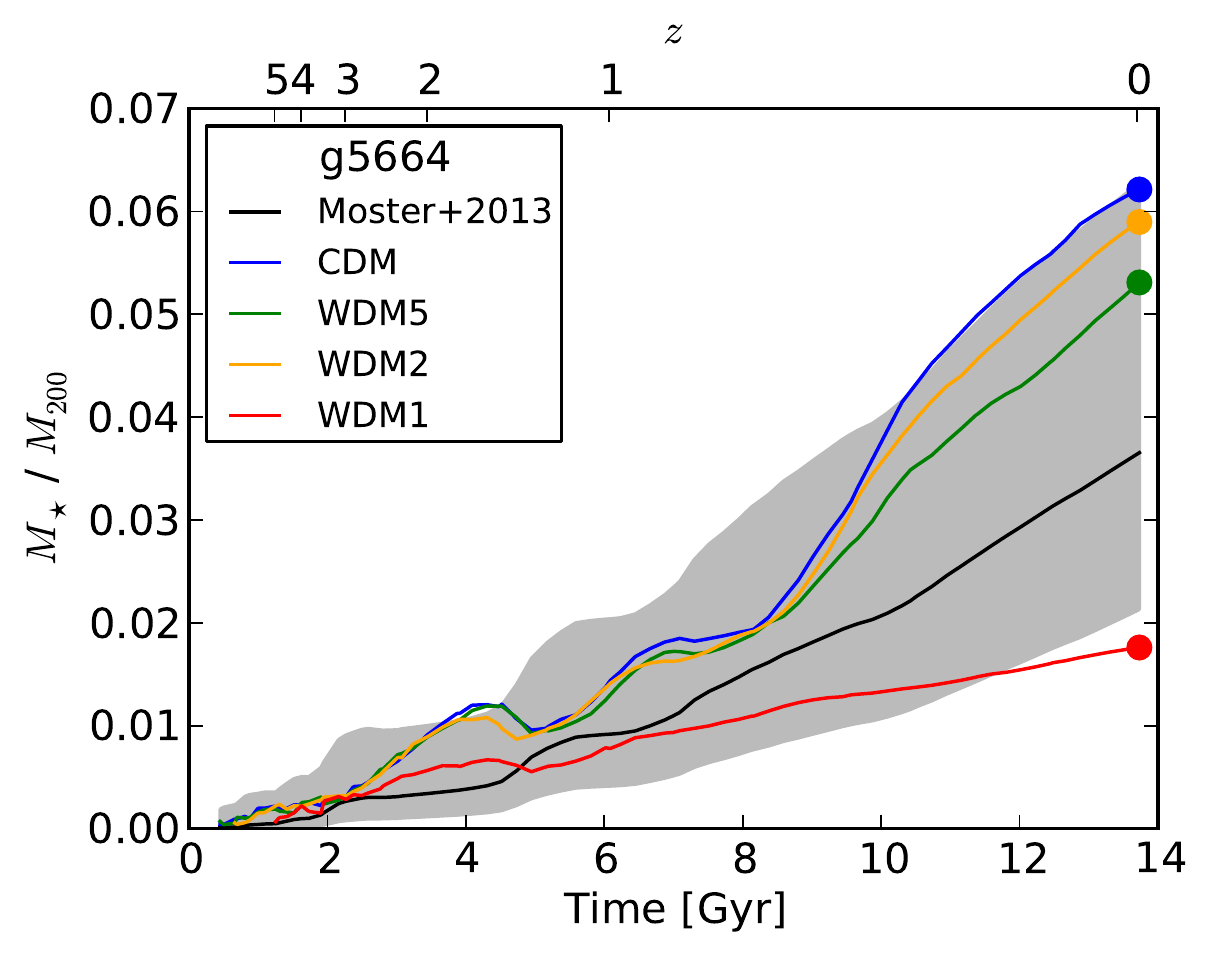}%
\includegraphics[width=.5\textwidth]{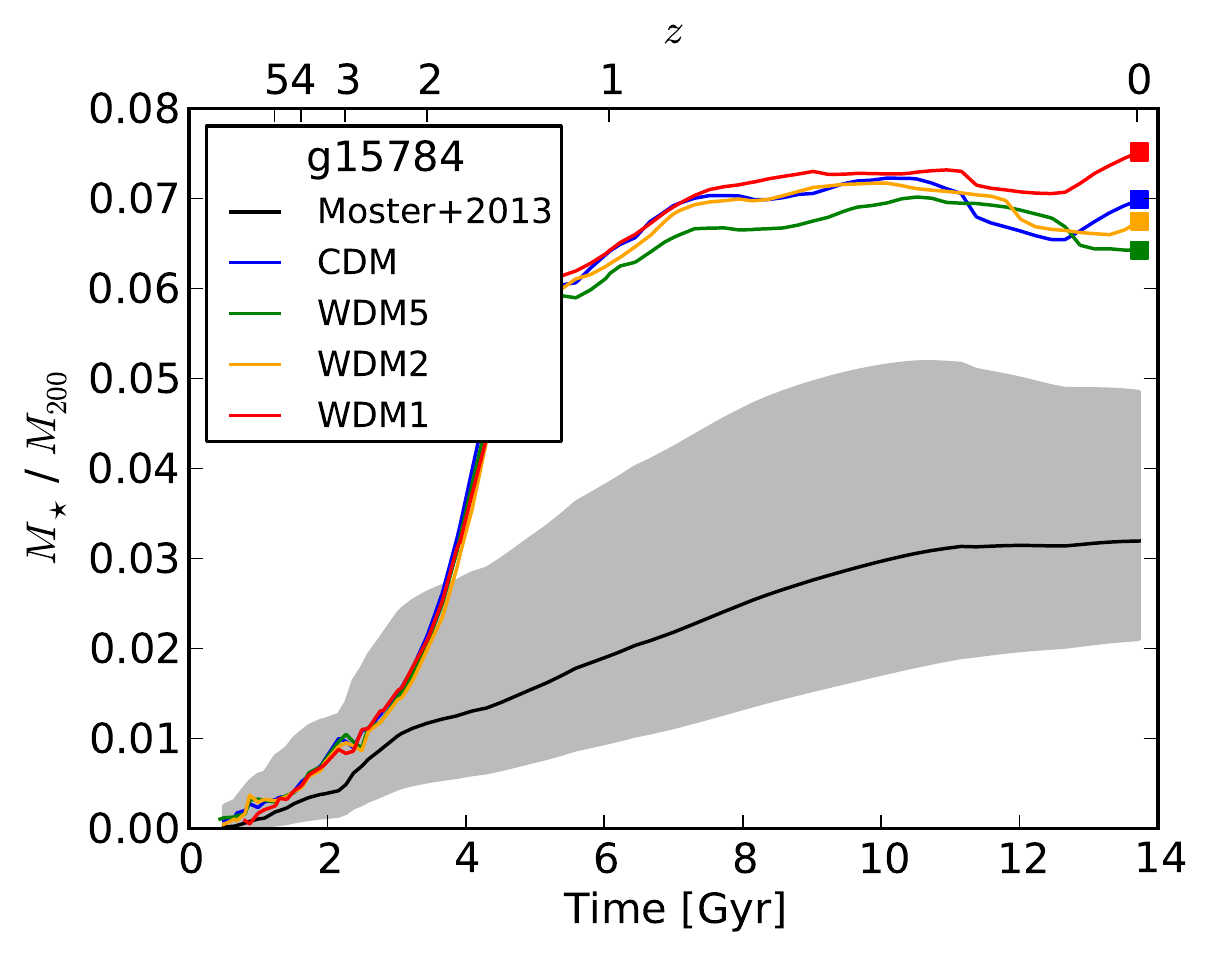}
\caption[$M_star$--$M_{\rm halo}$ relation]{
    Stellar mass--halo mass relation of the simulated galaxies.
    The upper left panel shows the stellar mass of all simulated galaxies as a
    function of total halo mass at $z=0$.
    The other panels show the evolution of the stellar mass--halo mass ratio
    as a function of cosmic time for each galaxy (upper right: g1536,
    lower left: g5664, lower right: g15784).
    The black lines visualize observational results obtained from abundance
    matching \citep{Moster2013} and the shaded area represents the corresponding
    $1\sigma$ scatter.
    While all implementations of g1536 and g5664 ($M_{\rm halo} < 10^{12} \msun$)
    roughly match observational results all simulations of g15784
    ($M_{\rm halo} > 10^{12} \msun$) produce too many stars.
}
\label{fig:moster}
\end{figure*}

The upper-left panel of Fig. \ref{fig:moster} shows the stellar mass of all galaxies as
a function of their halo mass at $z=0$.
All simulations of the g1536 galaxy (star symbols) {fall within the
expected variation in the relationship}.
While all DM models yield the same halo mass ($\approx6\times10^{11}\,\msun$),
there is a trend towards lower stellar masses with decreasing values for \mwdm.

{Each of the other three panels of Fig. \ref{fig:moster} show how the 
stellar mass to halo mass ratio varies as a function of time for each galaxy
individually.} The black
line shows the evolution based on the stellar mass predicted using the 
Moster relation with the evolution of the halo mass.  The grey region
represents the $1\,\sigma$ variation in this measurement.
The upper-right panel shows the results for all realizations of g1536.
It shows that the trend of reduced stellar content of the galaxy in \wdm 
starts at $z\approx1$.
From this panel, we also see that {every realization of} g1536 evolves 
consistently with the Moster relation at all times.

g5664 is denoted with {circles} in Fig. \ref{fig:moster}.
At $z=0$ (upper-left panel) the \cdm, \wdm{}5 and \wdm{}2 galaxies have the same halo
mass ($\approx4.5\times10^{11}\,\msun$) and almost the same stellar mass.
Their stellar mass is above the expected value but {remains marginally} consistent with the scatter.
The \wdm{}1 galaxy is slightly less massive in halo mass and has dramatically fewer stars.
Its stellar content {falls below the expected} Moster relation 
{variation} at $z=0$.
All simulations of g5664 evolve consistently with the Moster relation
until {$z<1$} (lower-left panel) but while the \cdm, \wdm{}5
and \wdm{}2 models {all sharply increase their stellar mass}, 
the \wdm{}1 model {forms far fewer stars from} $z\approx3$.

Looking at the other \wdm models, the clear trend of decreasing stellar mass 
with warmer DM models at $z\lta1$ in g1536 does not happen in g5664.
The \cdm model still forms the most stars, but the \wdm{}2 simulation forms
more stars than the \wdm{}5 model.

All simulations of the g15784 galaxy {have about the same halo mass} 
($\mtwo>10^{12}\,\msun$), but form {more} stars than are expected from
the Moster relation (squares in upper-left panel of Fig. \ref{fig:moster}).
The stellar mass--halo mass ratios of g15784 {evolve along
similar trajectories} for {all the initial conditions} (lower-right panel).
Their stellar {mass increases far above the expected variance} at $z=2$.
The {increase} is due to an extended SF event that
cannot be halted by any of the feedback processes that {are included in our 
simulations} \citep[for other examples, see][]{Kannan2013}.

The case of g15784 {illustrates} that {any effects that} \wdm 
may have on galaxy formation are {dwarfed} by the problem of 
overcooling in the mass regime of $\mtwo\gta10^{12}\,\msun$.
Thus, {for our study of the effect of \wdm on galaxy formation,
it is unnecessary for us to consider this galaxy} for further analysis.
The behaviour of g15784 also helps illustrate the need to use {\it realistic} galaxies
to study the effect of any variation of the cosmological
parameters, whether it be the properties of DM or dark energy.  Overcooling is a
common problem in simulations, and its effects can overshadow the effects of cosmology.

\subsubsection{Mass distribution}
\begin{figure*}
\includegraphics[width=\textwidth]{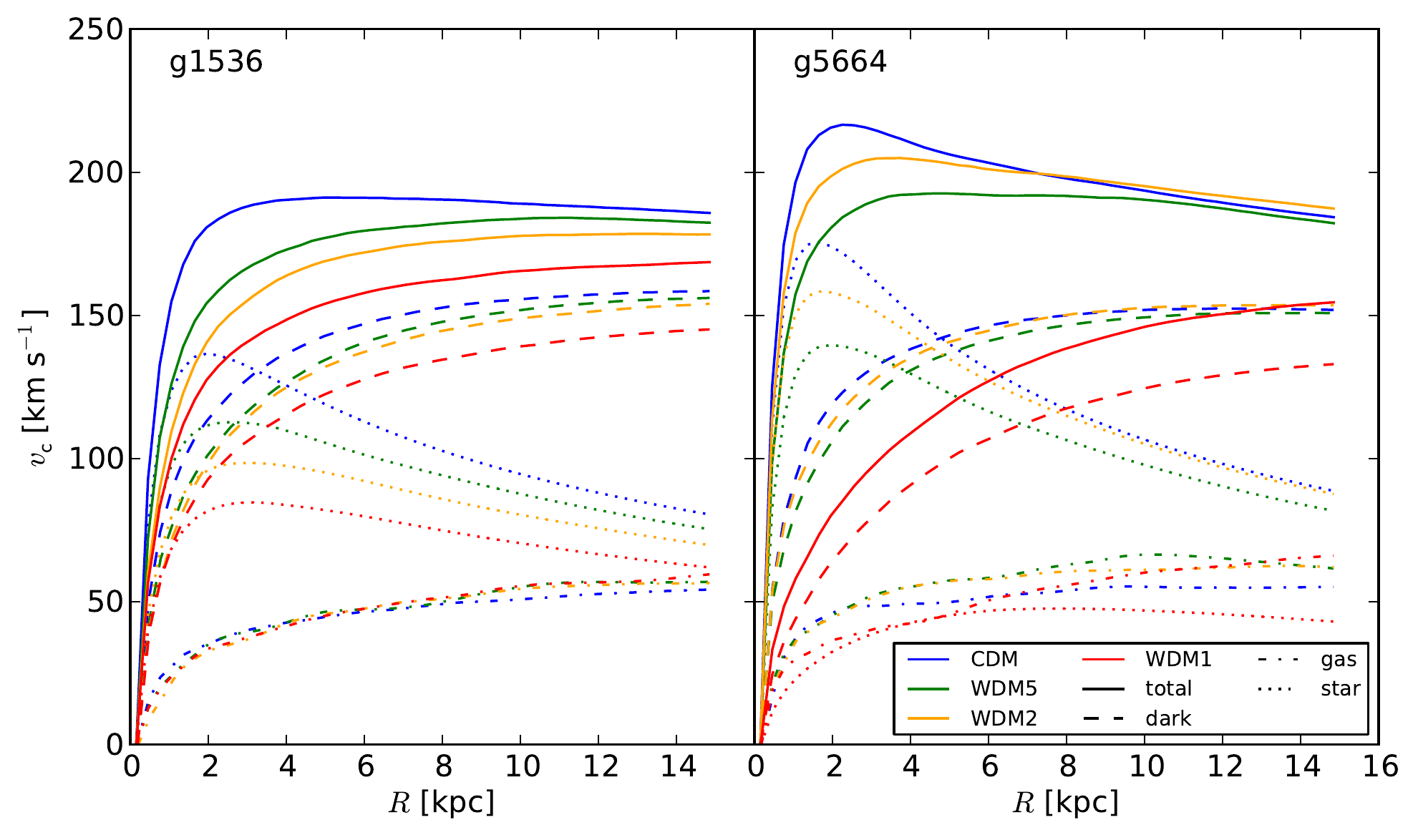}
\caption[Partial rotation curves]{
    Partial rotation curves of the hydrodynamic simulations of g1536 (left-hand panel) and
    g5664 (right-hand panel).
    The plot shows the circular velocity $v_{\rm c}=\sqrt{\mathrm GM\left(<R\right)/R}$
    as a function of $R$.
    The solid lines represent the total rotation curves, the dashed, dash-dotted and dotted
    lines the contributions of DM, gas and stellar mass, respectively.
}
\label{fig:rcs}
\end{figure*}
Fig. \ref{fig:rcs} shows circular velocities $v_{\rm c}=\sqrt{\mathrm GM\left(R\right)/R}$ as a
function of radius $R$ of the g1536 (left-hand panel) and g5664 (right-hand panel)
galaxies (solid lines).
The total rotation curve is split into the individual contributions from DM
(dashed), stellar
(dotted) and gas (dash-dotted) mass $M(<R)$.
All rotation curves for g1536 are {either} flat {or slowly rising}. 
{Warmer DM models (i.\,e. decreasing \mwdm) have} lower 
{maximum} rotation velocities and are more {slowly rising}.

The rotation curve for the \cdm model for g5664 has a central peak 
as we would expect based on its high rate of SF.
The central peak is less pronounced in the WDM simulations, and the {\wdm{}1 model
that formed too few stars has a slowly rising rotation curve.  Like for the
SF, the rotation curves of the \wdm{}2 and \wdm{}5 are in the 
opposite order in g5664 from what they are in g1536.}

For both galaxies, the contributions {to the rotation curves} from DM
and gas are quite similar {between} the different DM models.
The {largest difference in contribution to the rotation curve comes from}
the stars {in the central} 5\,kpc.
Thus, the difference in stellar masses {apparent in Fig. \ref{fig:moster}
results from excess SF in} the centre of the galaxies, rather
than stars accreted from satellites. 

\begin{figure*}
\includegraphics[width=\textwidth]{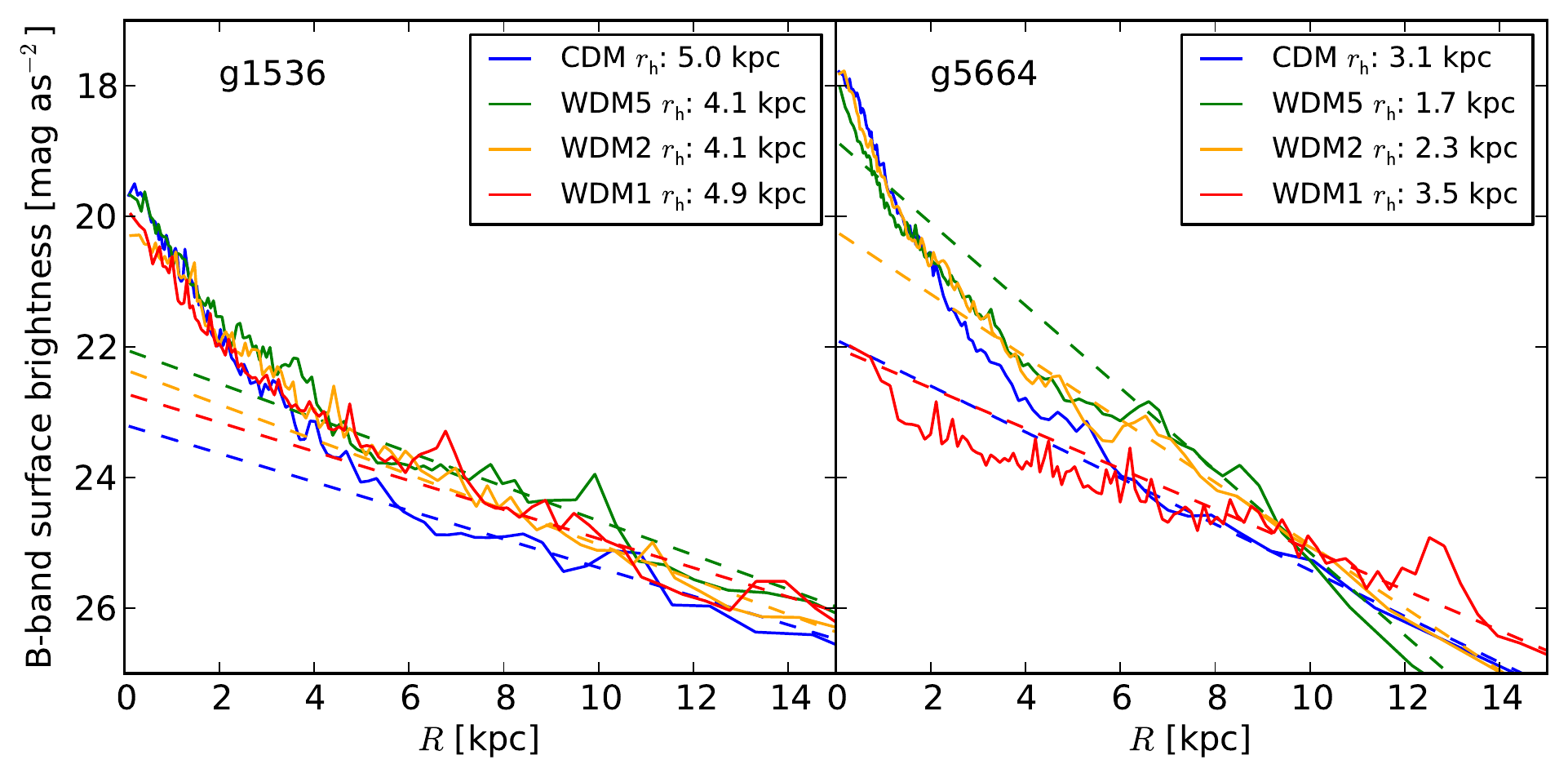}
\caption[Surface brightness profiles]{
    $B$-band surface brightness profiles of g1536 (left-hand panel) and g5664 (right-hand panel).
    The solid lines show the inferred surface brightness profiles, the dashed lines
    show the result of exponential fits to the profiles for $R>{\rm 7\,kpc}$.
    Except for the WDM model of g5664, all galaxies feature a central bulge.
    It is less pronounced for warmer DM models in the case of g1536.
}
\label{fig:sbprofs}
\end{figure*}

{Another, more directly observable, way to look at the mass distribution
in our simulated galaxies is the} $B$-band surface brightness profiles shown 
in Fig. \ref{fig:sbprofs}.
Again, the left-hand panel shows data for g1536 and the right-hand panel for g5664.
Except for the \wdm{}1 simulation of g5664, all galaxies feature an
exponential disc with a central bulge.
The profiles  have been fitted {(dashed lines)} with an exponential outside 
of $7{\rm\,kpc}$. The resulting scale-lengths $r_{\rm h}$ are given in the 
legends.

{The two different galaxies show a range of scale-lengths.  The g5664
models are all shorter than the g1536 realizations.}
The \wdm{}5 model of g5664 {reports a particularly} short scale-length
because its surface brightness profile has a flat
feature at approximately $6{\rm\,kpc}${, and then a sharp break outside
that}.  Since the exponential was only fitted outside $7{\rm\,kpc}$ it 
results in a steep scale length.

At $z=0$, {the time at which the profiles are shown,} the galaxy g5664 is 
{in the process of} accreting a satellite.
The exact timing of the merger {varies between} the \wdm models.
The flat feature in the $B$-band surface brightness profile in the 
\wdm{}5 and \wdm{}2 models {results from the} tidal interaction 
with the satellite, while the `bump' at approximately $13{\rm\,kpc}$ in 
the \wdm{}1 model is the satellite itself.

The surface brightness serves as a tracer of mass, {so it is no surprise
that the surface brightness profiles} roughly follow the same trend with \mwdm as the 
total stellar mass {and rotation curves} of the halo: {models with less 
stellar mass have less bright centres and less centrally peaked rotation
curves.}  This supports the conclusion drawn from the 
rotation curves that the major differences in stellar abundance among the 
different DM models occur in the galaxies' centres.

\subsubsection{Mass accretion evolution}
{Since the SFH is known to have a significant impact on the
evolution of the galaxies, we investigate how gas is supplied to the disc
out of which those stars form.}
\begin{figure*}
\includegraphics[width=\textwidth]{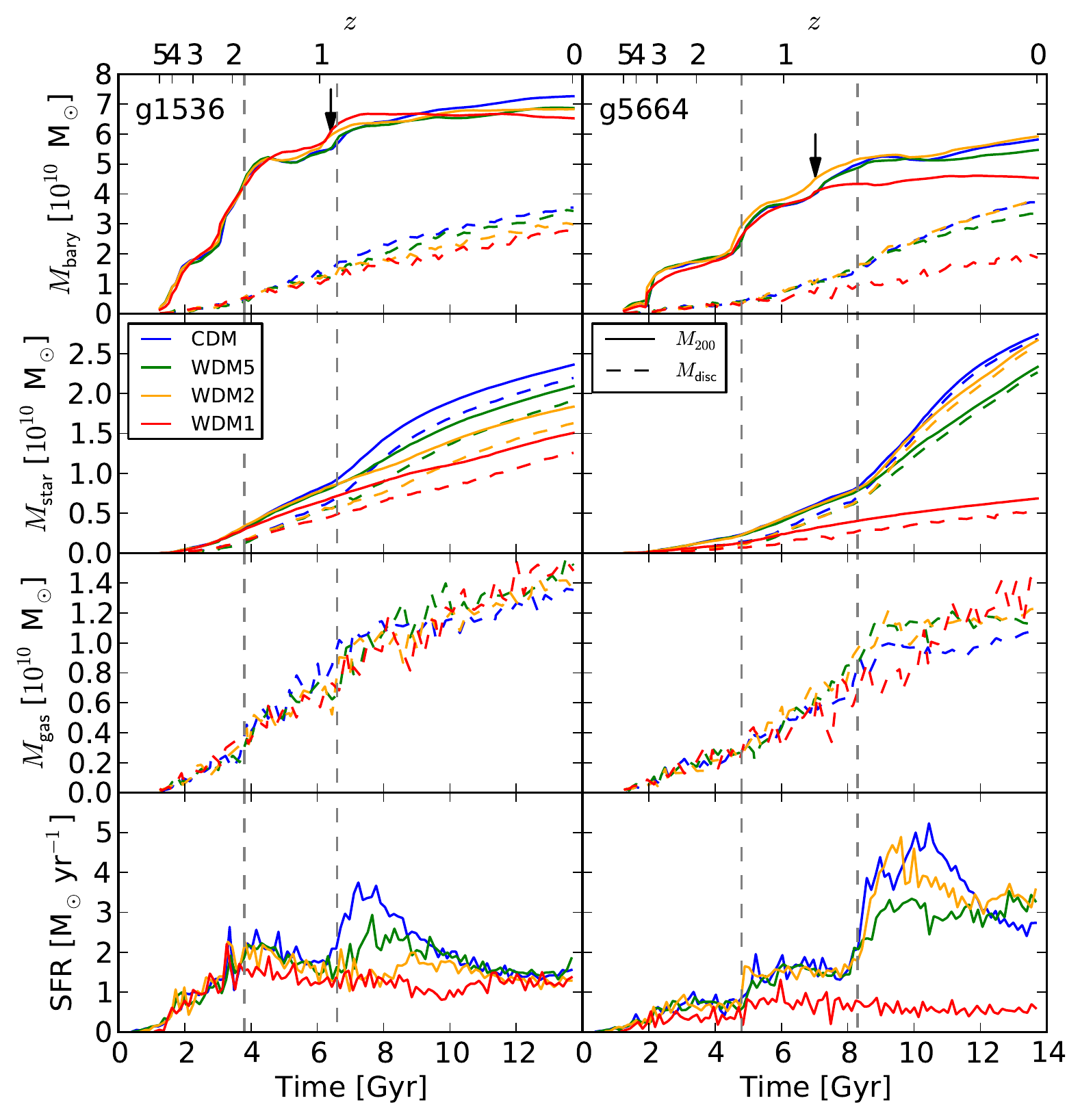}
\caption[Baryonic mass evolution and star formation history]{
    Accretion histories of baryons and SFH for g1536 and g5664.
    The plots show the evolution of baryonic (upper row), stellar (second row) and
    gas (third row) mass of the simulated galaxies' main progenitor.
    The bottom row shows the SFH of the galaxy.
    The left-/right-hand panels represent the results for g1536/g5664.
    Solid lines represent all mass within the virial radius and the dashed lines
    represent all mass that is closer than 3 comoving kpc to the plane of the disc and
    20 comoving kpc to the disc's rotational axis.
    The gas mass of the entire halo is not shown.
    Sudden increases in baryonic mass (upper panels) indicate merging events with satellite
    galaxies.
    The vertical dashed lines are a guide for the eye and indicate those points in time at
    {which enhanced SF occurs in the CDM models.
    The arrows indicate the time of the last significant merger.
    In each of the DM models, this merger happens within $\pm200{\rm\,Myr}$
    around the time, marked by the arrow.}
}
\label{fig:massaccretion}
\end{figure*}
Fig. \ref{fig:massaccretion} shows the cumulative mass evolution of the 
{various components of the} baryonic 
matter along with the SFH (bottom row).
The left- and right-hand columns show results for g1536 and g5664, respectively.
The evolution of the total mass of baryons is shown in the top row, 
stellar mass evolution is shown in the second row and
gas {evolution is shown in the} third row.  {Solid lines show the
mass contained within the virial radius, while the dashed lines show the 
mass} in a disc {that is} 6 comoving kpc thick with a radius of
20 comoving kpc (dashed lines).
The disc volume was aligned based on the angular momentum of gas
{inside 3 kpc} of the respective halo.
The bottom panel shows the SFH of all stars bound in the $z=0$ halo.

{Merging events (\ie the time at which a satellite enters the virial radius)
are apparent as sudden increases in} the total baryonic 
mass.
{The arrows in the top panel of Fig. \ref{fig:massaccretion} indicate an average
time when the last significant satellite crossed the virial radius in the
different DM models.\footnote{The arrow indicates the average between the earliest
and the latest merging time.
These extreme times are $<200{\rm\,Myr}$ from the time indicated by the arrow}
The satellite accreted in the \wdm{}1 model of the galaxy g5664 has about half the mass
of the corresponding satellite that is accreted in the other DM models.
The lower satellite mass results in the reduced SF shown in the \wdm{}1 model
of g5664 in Fig. \ref{fig:moster}.
The vertical dashed lines in Fig. \ref{fig:massaccretion}
indicate the onset of star `bursts' in the two \cdm models.
}

The onset of the most prominent SF feature in
g1536 ($t\approx 6.5{\rm\,Gyr}$, second vertical dashed line) {occurs just after
this last merging event.}
After that merging event, the stellar mass {evolution} of the {various models separate
{such that the increase in SF is larger for more massive particle candidates.}
{Over the course of the next} $4{\rm\,Gyr}$, the SF rates {in the models
that saw the greatest increase in SF gradually decrease back} to the same 
{SF rate as the models that saw no increase in SF.}

The top panel shows that {all the models gain baryonic mass at nearly identical rates
with the slight exception that} the \wdm{}1 halo {has more baryons before and until
2 Gyr} after the merger {at which point the baryon content of the other models surpasses
\wdm{}1}.
Thus, the reduced stellar mass in the \wdm{}1 model cannot be explained {as
a simple, straightforward result of} less
baryons {leading to less SF.  Instead, the cause of the SF event must
be more subtle for g1536.}  The {evolution of the} baryonic content of the disc of g1536 
{shows exactly the same trend as SF and hence stellar mass.
Partly, this is due to the fact that more 
SF leaves more stars in the disc, but the evolution of the disc gas mass shown in the
third panel shows that while all the discs contain about the same amount of mass (note the
different scales for each panel), the \cdm
model generally contains more until well after the merger when the SF is more 
efficient, so SF consumes the gas and the total gas content drops below the 
other models.}

{Contrary to g1536,} the \wdm{}1 model of g5664 {shows a large discrepancy in
total baryonic mass from the other models following a merger at $z\sim1$.
The discrepancy leads directly to much less SF.} 
{Regarding the} baryonic mass {content} of the disc, {the different \mwdm
models for} g5664 do not {line up sequentially according to \wdm candidate 
mass. However,} as in g1536, the stellar mass of the disc follows
the baryonic mass of the disc.
Thus, the SFH most {clearly correlates with the mass} of baryons in the
disc of the galaxy, {which is the result of how efficiently} gas is accreted on to the disc.
Once the gas is accreted to the disc it {reaches densities where it can form stars}.
The correlation between stellar and baryonic mass in the disc is less pronounced but also
evident after the merging events {preceding} the first vertical dashed line 
in Fig. \ref{fig:massaccretion} for both galaxies.

\subsection{The effects of merging}
{Significant differences in SF (indicated by the vertical dashed lines in
Fig. \ref{fig:massaccretion}) occur just after merging events (indicated by increases in
total baryonic mass in Fig. \ref{fig:massaccretion}).}
{For both galaxies studied, we saw that} the efficiency
of SF correlates to the baryonic mass of the disc.
{We now take a closer look at how the differences in the \wdm models
can have different effects on the disc.  Each model has nearly the same 
merger history, but because of the changes to the initial power spectrum
inherent in our \wdm model, the satellites that merge have slightly
different properties.  We find that the way these satellites evolve changes
the metal content of the halo and the magnitude of disc instabilities
excited in the disc, both of which lead to changes in the SFH.}

This difference in the satellites' properties is not surprising.
The effect of WDM on structure formation becomes stronger when the
halo mass approaches the cut-off scale in the power spectrum
\citep[\eg][]{Bode2001,Schneider2012,Kang2013}.
The satellites we discuss in this section have a mass of
$\approx 10^{10} \Msun$ and hence they are more sensitive
to the effects of the WDM component.

\subsubsection{Disc instabilities}

Dynamical instabilities in discs collapse gas to higher densities.
Discs collapse due to instability when the potential determined by their
surface density ($\Sigma$) is too great for their velocity dispersion
and their epicyclic frequency ($\kappa$) to support.  Thus, as discs grow
in mass, they become more prone to unstable collapse.  Such collapse can
create high density regions in which stars can form more efficiently.
Interactions with satellites can also drive instabilities in the disc
\citep{Mihos1996}.

\begin{figure*}
\includegraphics[width=\textwidth]{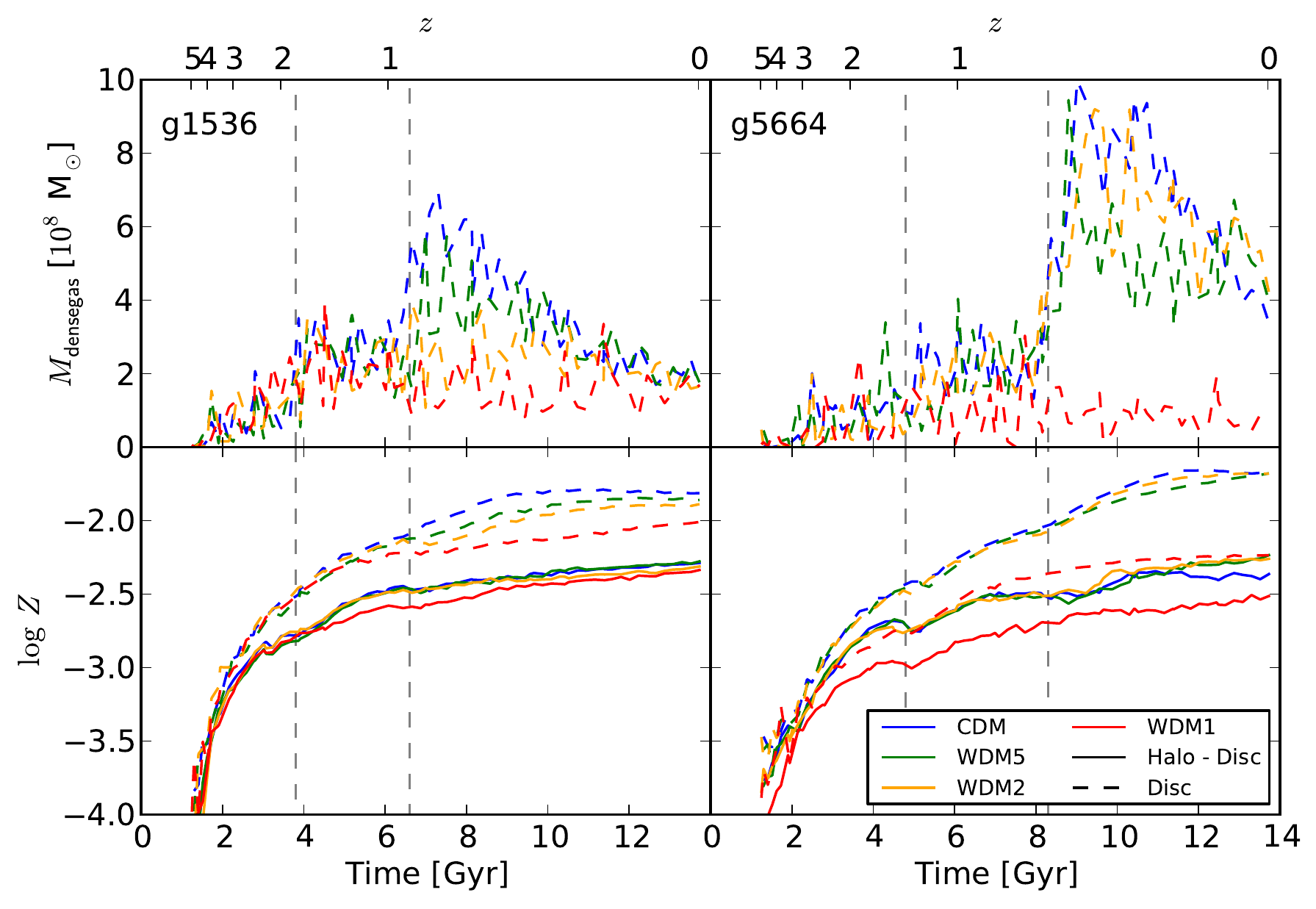}
\caption[Metallicity evolution]{
    Evolution of the dense gas mass (top row) and metallicity (bottom row)
    of the gas in g1536 (left) and g5664 (right).
    The dashed lines show the dense gas mass and metallicity of the gas in the disc
    (6 comoving kpc thick and 20 comoving kpc radius) and the solid lines in the
    bottom panels show
    the metallicity for all the gas in the halo without the disc.
    The vertical dashed lines are the same as in Fig. \ref{fig:massaccretion},
    \ie they indicate the onset of major star formation effects.
}
\label{fig:feh}
\end{figure*}

The top panels of Fig. \ref{fig:feh} show that at the time of the 
SF increase associated with the second merger (right grey
dashed line) there is a dramatic rise in the quantity of dense ($n>n_{\rm th}$)
gas in the \cdm simulation. There is progressively less dense gas in warmer
DM candidates in the case of g1536 (left).
The increase in dense gas suggests that the
tidal interaction with the accreting satellite in the \cdm model
is effective at driving instabilities to increase SF.
In the case of g5664 (right) the coincidence of the onset of SF
and the excess of dense gas in the disc is also evident.
However, we do not see the trend with warmer DM which is consistent
with what we saw in stellar mass evolution and SFH
(\cf Fig. \ref{fig:massaccretion}).
Note that the delay between the merging event and the corresponding star formation
feature is much longer for g5664 compared to g1536.
This is due to the satellites' orbits.
While the satellite of g1536 falls straight into the centre of the host galaxy
the satellite of g5664 has a more tangential orbit and approaches the host's centre
much slower.
Thus, its effects on the host's disc are delayed more.

\subsubsection{Gas halo metal enrichment}
The bottom panels of Fig. \ref{fig:feh} show the evolution of the metallicity in the
gas disc (dashed lines) and the gas halo outside the disc (solid lines)
for all the \wdm models in g1536 (left-hand panel) and g5664 (right-hand panel).
Elements with atomic masses higher than hydrogen and helium are produced through
fusion reactions inside stars, with the most massive elements forming
in massive stars that have evolved off the stellar main sequence and 
started fusing elements heavier than hydrogen and helium.  When these
stars explode as SNe, the metals leave the star as part of a
hot ($>10^6$ K) gas phase that has enough energy to leave the disc potential.
Some of the metals mix into the interstellar medium gas that comprises
the disc, increasing the disc gas metallicity, as shown in Fig. \ref{fig:feh}.
The rest of the metals leave the disc to create and enrich the hot gas halo.

Stars formed in satellites also create metals.  Since the satellites in 
our models have lower mass than the main disc, their gravitational potential
is shallower, so it is easier for the hot gas to leave the satellites.

The most significant difference in metal content of the gaseous haloes in
the g1536 models happens after the first grey dashed line at
4.2 Gyr.  We see in the simulations that there are three minor satellites
accreting on to the disc at this time.  The three satellites have reasonably
similar properties in the \cdm, \wdm{}5 and \wdm{}2 models, but have many
fewer stars and less dense gas in the \wdm{}1 model.  The satellites that
have formed a few stars when they enter the halo bring cool, metal poor
gas into what is already a metal enriched main halo.  The cool gas is
not ram-pressure stripped initially because the depth of the satellite potential
well is too deep.  As the satellites reach the pericentre of their orbit,
they are tidally stirred \citep{Mayer2001} and develop dense gas regions.
These dense regions form stars and blow out hot, metal-enriched gas
much more efficiently than the satellites had previously.  The difference
in halo metallicity between the \wdm{}1 model, in which the satellites
were too tenuous to form stars, and the other models in which stars form
is clear in Fig. \ref{fig:feh}.
This low efficiency of SF in the WDM1 satellites is 
due to a combination of their later formation time \citep[\eg][]{Macci`o2010}
and due to their reduced central concentration \citep[\eg][]{Schneider2012}.

For the g5664 galaxy, the situation is similar.
The halo metallicity in the \wdm{}1 model is lower than that of the other DM models.
However, here this difference occurs before the first major onset of SF,
\ie even at the highest redshifts at which we were able to identify
the galaxy's main progenitor.
This indicates that SF in the \wdm{}1 model of g5664 stays behind compared
to colder DM models due to delayed halo formation in warmer DM.

The difference in halo metallicity between the \wdm{}1 and the other DM models
decreases at late times since the metal enriched gas can escape from the disc
in the \wdm{}1 model more easily due to its shallower potential well.

The result of the increased metallicity in the gas halo is that the cooling
rate increases. Thus, gas cools quicker out of the halo 
on to the disc. As we have seen, more gas in the disc leads to more SF.

\subsection{WDM versus stellar feedback}
\label{sec:reducedfeedback}
\S \ref{sec:wdmsims} {illustrated how} the nature of DM can have 
an effect on the morphology of disc galaxies.
It is interesting {to compare} these effects {with} the influence of
baryonic physics on the process of galaxy formation.
In order to test this, a second series of simulations {using} the most extreme \wdm model
(\wdm{}1) was carried out with {a variety of} ESF 
efficiencies, \cesf, from none to {the fiducial value}.
The evolution of the stellar mass--halo mass ratio for these simulations is presented in
Fig. \ref{fig:evmosteresf}.  {The results reflect what was found in
\citet{Stinson2013}:  ESF effectively delays SF.
Even using the most extreme \wdm model, too many stars form early with the 
slightest reduction of \cesf.}
Simulations with $\cesf\lta0.075$ form too many stars at high $z$.
{One interesting trend is apparent: The simulations without ESF
form less stars by $z=0$ than some simulations with a
small amount of ESF.}
{The final stellar mass is lower without ESF because gas collapses to higher
densities without ESF, which causes more intense starbursts that }
drive stronger outflows.
These outflows remove large amounts of gas leading to a lack of SF fuel at later times
which leads to the suppressed SF at $z\lta2$.
This effect has been described by \citet{Stinson2013} in \cdm and applies to
galaxy formation in \wdm models as well.
The fact that it occurs in the most extreme case of the explored \wdm models shows that
the effects of \wdm are {minimal compared to} baryonic feedback processes.

\begin{figure}
\includegraphics[width=\columnwidth]{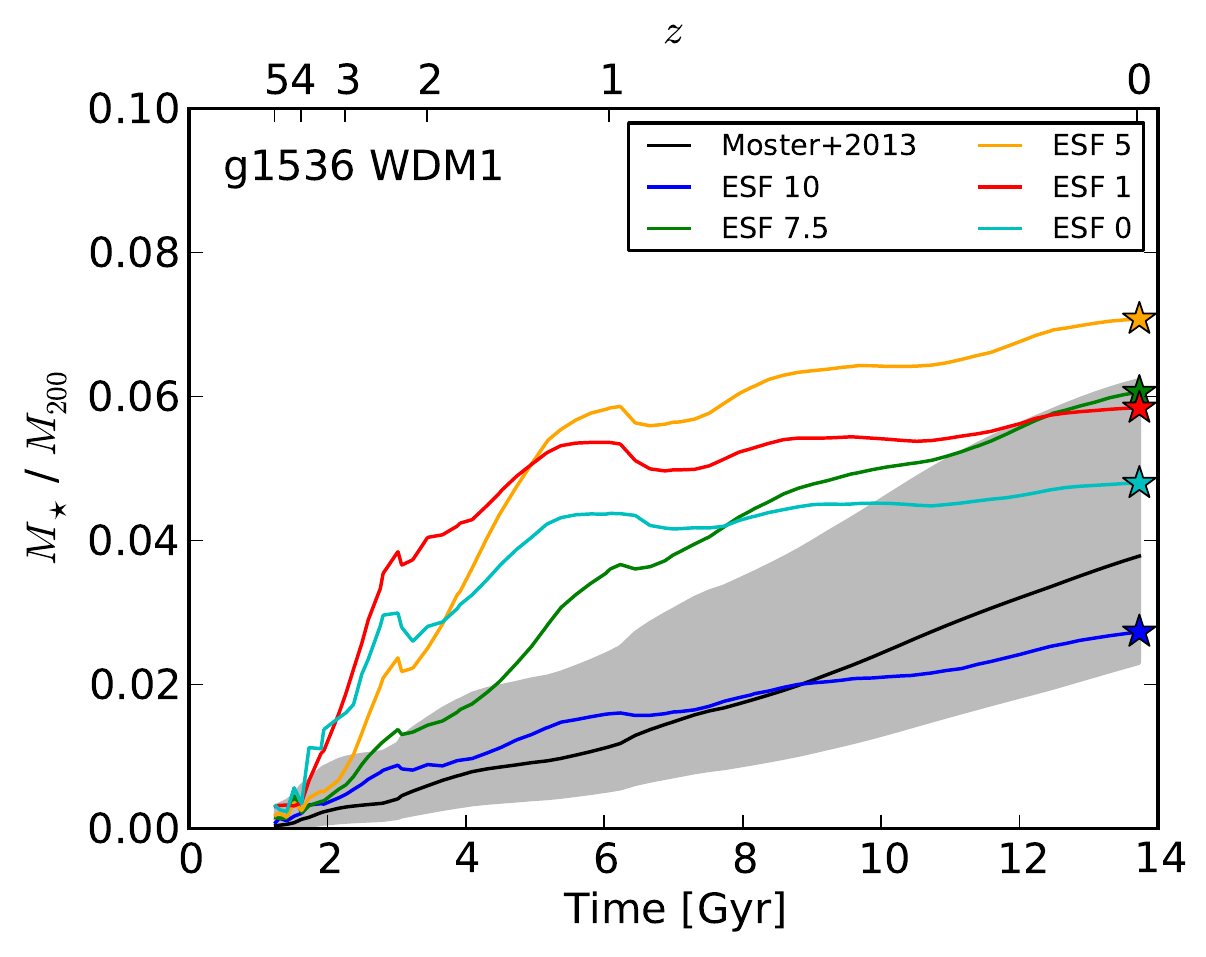}
\caption{
    Evolution of the stellar mass--halo mass ratio for simulations of g1536 WDM1
    with different ESF efficiencies.
}
\label{fig:evmosteresf}
\end{figure}

\section{Conclusions}
\label{sec:conclusions}

We carried out a series of hydrodynamical cosmological zoom simulations of the formation
of three different galaxies.
Each galaxy was simulated in four different DM models ranging from 
the standard CDM scenario to \wdm models
with a DM candidate mass as low as $\mwdm=1{\rm\,keV}$.

Galaxy formation proceeded in a similar fashion regardless of the \wdm
model used. Overall WDM reduces the SF rate, which results in a marginally
lower total stellar mass at $z=0$ and in less centrally concentrated stellar profiles. 

We find that these differences are mainly due to the different properties
of the few satellites merging with the central object.
In the CDM model satellite interactions
triggered disc instabilities that drove gas to the centre where it 
greatly increased the rate of SF and consequently
raised the rotation curve in the centre. In some cases, the increase
in SF was simply due to the creation of dense gas in the
tidal interaction with the satellite. In other cases, the increase
in SF could be traced to metals ejected from the satellites into
the gas surrounding the main disc. The metal-enriched halo proceeded to cool 
faster on to the disc, which increased SF. 

While the CDM and the WDM galaxies share the same merger histories 
the properties of the incoming satellites differed from model to model, most
significantly between the extremes of our simulations, the coldest, CDM, and warmest, WDM1, models.
Due to the later formation time and the reduced DM concentration of the WDM1 satellites
that are predicted in several $N$-body studies \citep[\eg][]{Schneider2012}
the satellites in the warmest model were least efficient in retaining their gas and forming stars.
As a consequence they had the least (hydro)dynamical impact on the central object
during the merging phase, and did not trigger any bursts of SF.

The reduced impact of satellites in the WDM1 model is mostly notable 
in the galactic rotation curves, which are indeed rising much slower
than in the \cdm simulations.

However, this effect could be matched with minor variations of the stellar feedback.
The series of tests simulated with a variation of the ESF
(feedback from massive stars before they explode as SN) showed
that the feedback had a stronger effect than the difference between CDM and the warmest,
1\,keV WDM1 model.

A clear signature 
of the effects of a \wdm component was only seen in two 
out of the three simulated galaxies. The effects only appeared in galaxies that had realistic
properties like stellar mass--halo mass ratio and a realistic
rotation curve. The third galaxy (due to its higher mass)
was too overcooled to produce any of the observed
galaxy scaling relations. In this third galaxy, the
effects of WDM were erased by the unrealistically high SFE.
These results again stress the need for \emph{realistic} 
hydrodynamical simulations to assess the effect of cosmology on galaxy properties
\citep[\eg][]{Casarini2011,vanDaalen2011}.

Finally, the only \wdm model to show
a clear impact on galaxy properties was the WDM1 model. 
This model falls outside the current constraints on WDM particle mass.
If the new limits by \citet{Viel2013} (\wdm candidate masses below $3.3\,{\rm keV}$ 
ruled out at a confidence level of $2\sigma$)
can be confirmed, our simulations show that the effect of WDM on disc galaxy formation 
is minimal (if not totally absent) especially when compared to 
processes such as stellar feedback.

\section*{Acknowledgements}
The analysis was performed using the {\sc pynbody} package
(\texttt{http://code.google.com/p/pynbody}), which had key contributions from 
Andrew Pontzen and Rok Ro\v{s}kar in addition to the authors.
The simulations were performed on the \textsc{theo} cluster of  the
Max-Planck-Institut f\"ur Astronomie at the Rechenzentrum in Garching;
the clusters hosted on \textsc{sharcnet}, part of ComputeCanada; the Universe
cluster that is part of the \textsc{cosmos} Consortium at Cambridge, UK; the
\textsc{hpcavf} cluster at the University of Central Lancashire;
and the Milky Way supercomputer,
funded by the Deutsche Forschungsgemeinschaft (DFG) through Collaborative Research
Center (SFB 881) `The Milky Way System'
(subproject Z2), hosted and co-funded by the J\"ulich Supercomputing
Center (JSC). 
We greatly appreciate
the contributions of all these computing allocations.
AVM and GSS acknowledge support through the  Sonderforschungsbereich SFB 881
`The Milky Way System'
(subproject A1) of the German Research Foundation (DFG).
CB  acknowledges Max- Planck-Institut f\"ur Astronomie for its hospitality
and financial support through the  Sonderforschungsbereich SFB 881 `The Milky Way System'
(subproject A1) of the German Research Foundation (DFG).
HMPC and JW gratefully acknowledge the support of NSERC.  HMPC also appreciates the
support he received from CIfAR.

\clearpage

\end{document}